\documentclass{aastex}
\usepackage{emulateapj5}
 
\def\kms{km~s$^{-1}$}

\def\ga{\mathrel{\hbox{\rlap{\hbox{\lower4pt\hbox{$\sim$}}}\hbox{$>$}}}}
\def\la{\mathrel{\hbox{\rlap{\hbox{\lower4pt\hbox{$\sim$}}}\hbox{$<$}}}}
 
\shorttitle{}
 
\shortauthors{S.\ Stanimirovi\'{c} et al.}
 
\received{2003 February 17}
\begin{document}
 
\title{Detection of OH absorption against PSR B1849+00}
 
\author{Sne\v{z}ana Stanimirovi\'{c}}
\affil{Radio Astronomy Lab, UC Berkeley, 601 Campbell Hall, Berkeley, CA 94720}
\email{sstanimi@astro.berkeley.edu}
\author{Joel M. Weisberg}
\affil{Department of Physics and Astronomy, Carleton College, Northfield,
MN 55057}
\author{John M. Dickey}
\affil{Department of Astronomy, University of Minnesota, 116 Church St. SE,
Minneapolis, MN 55455}
\author{Anton de la Fuente, Kathryn Devine, Abigail Hedden}
\affil{Department of Physics and Astronomy, Carleton College, Northfield,
MN 55057}
\author{Stuart B. Anderson}
\affil{Department of Astronomy, MS 18-34, California Institute of
Technology, Pasadena, CA 91125}
\begin{abstract}

We have searched for OH absorption against seven pulsars using the Arecibo
telescope. In both OH mainlines (at 1665 and 1667 MHz), deep and narrow 
absorption features were detected toward PSR B1849+00. 
In addition, we have detected several absorption
and emission features against B33.6+0.1, a nearby supernova remnant
(SNR). The most interesting result of this study is that a pencil-sharp
absorption sample against the PSR differs greatly from the large-angle
absorption sample observed against the SNR. If both the PSR and the SNR probe the
same molecular cloud then this finding has important implications for
absorption studies of the molecular medium, as it shows that the statistics of
absorbing OH depends on the size of the background source.
We also show that the OH absorption against the PSR most likely originates
from a small ($<30$ arcsec) and dense ($>10^{5}$ cm$^{-3}$) molecular
clump. 

\end{abstract}

\keywords{line: profiles --- pulsars: individual (B1849+00) --- supernovae:
individual (B33.6+0.1) --- ISM: molecules --- ISM: structure}

\section{Introduction}

Much information concerning the interstellar medium (ISM) has been 
gleaned from studies of the absorption of pulsar signals by neutral 
hydrogen at $\lambda\sim 21$ cm. Kinematic distances derived from 
absorption spectra, in combination with pulsar dispersion measures, have 
determined the electron density throughout the plane of the Galaxy 
\citep{Weisberg95}.  
Combination of absorption and emission spectra along the same line-of-sight
allows us to determine optical depth and 
inferred spin temperature of the cold neutral gas in front of the 
pulsar \citep{Weisberg79,Koribalski95}.
Particularly interesting application of pulsar absorption studies 
is by \cite{Frail94} who found temporal 
variations in pulsar absorption spectra, suggesting structure in the
absorbing HI clouds at the  tens--of--AU scale 
(the tiny-scale atomic structure or TSAS).  
The origin and properties of TSAS are still a great puzzle.
Observations suggest that these features have pressure $>300$
times greater than the standard pressure of the cold neutral medium, being
far from in a pressure equilibrium with the surrounding medium, unless
they posses an extremely extended geometry \citep{Heiles97}.

Another interesting issue was addressed by \cite{Dickey81} who studied 
the existence of very small scale structure in the HI optical depth
distribution. They compared abundance of HI absorption lines in various
optical depth intervals measured against both continuum
extragalactic sources, with typical angular sizes of 1$''$ to 3$'$, and against pulsars. 
As both sets of sources showed similar abundance of HI absorption lines
they concluded that the angular size of background sources does not influence
the optical depth measurements in HI, at least over the angular size range 
0.001$''$ to 3$'$.

As pointed out by \cite{Dickey81}, pulsars are particularly powerful
background sources for emission-absorption studies
because the solid angle subtended by their continuum emission is extremely
small.  Although scattering in the ISM broadens their apparent size, 
the effective diameter
of the line-of-sight volume is still on AU scales.  This means that absorption
spectra against pulsars probe needle-thin samples of the ISM. 
Another great advantage which pulsars offer as background sources for such
studies, is that they turn on and off, thus allowing the absorption and
emission spectra to be measured without moving the telescope.  This
eliminates the errors introduced in the absorption spectrum by small scale
spatial variations in the emission.

Motivated by pulsars' unique capabilities for studying the ISM, and in an
effort to extend their use to molecular medium, we measured the absorption 
spectra of several pulsars at the wavelength of the hydroxyl radical (OH), 
$\lambda\sim 18$ cm.  The only previous attempts to determine 
pulsar OH absorption spectra were by \cite{Galt74}, who did not 
detect absorption in the spectrum of PSR B0329+54, and by \cite{Slysh72}, who 
observed B0329+54, B0450--18, B0525+21, B0740--28, B1642--03, and B1749--28; 
and detected OH absorption at 1667 MHz toward the latter one.
As pulsar emission is weak at 18 cm OH absorption measurements
are quite difficult and require a very sensitive instrument.
We used the Arecibo telescope and detected OH absorption 
against one of our sources -- PSR B1849+00.
The line-of-sight toward B1849+00 is particularly interesting as it passes
right through the Galactic plane and is very close to a nearby supernova
remnant (SNR) G33.6+0.1.
In this paper we focus on a detailed comparison between OH absorption
spectra toward B1849+00 and G33.6+0.1.

This paper is organized as follows.
We start in Section~\ref{s:background} with a brief background on 
the two objects of interest in the paper, B1849+00 and G33.6+0.1.
Section~\ref{s:observations} describes our Arecibo OH observations 
and data processing. To learn more about the large-scale distribution 
of molecular gas and compare it with OH we used the
Steward Observatory's 12-m telescope to map a 25$'\times$ 25$'$ region around
G33.6+0.1 in $^{12}$CO(1-0); these observations
are also summarized in Section~\ref{s:observations}.
Detected OH absorption spectra toward B1849+00 and G33.6+0.1 are presented
and analyzed in Section~\ref{s:psr-only} and Section~\ref{s:snr-only}.
The $^{12}$CO(1-0) distribution is presented in Section~\ref{s:CO}.
In Section~\ref{s:small-scale-structure} we discuss 
the main result of this paper -- the great difference between 
OH absorption spectra measured against the continuum emission 
from B1849+00 and G33.6+0.1 separately along the same line-of-sight.
We present two geometrical scenarios to explain this difference and
investigate several possible implications.
We conclude in Section~\ref{s:summary}.

\section{Background on PSR B1849+00 and SNR G33.6+0.1}
\label{s:background}

PSR B1849+00 is a low-latitude, $(l,b) = (33.5,0.0)$, 
long-period (2.18 s) pulsar discovered by \cite{Clifton86}. 
What is particularly interesting about this object is its very high 
dispersion measure of $(680\pm60)$ cm$^{-3}$ pc \citep{Clifton88}.
HI absorption measurements towards B1849+00
were obtained by \cite{Clifton88} 
who established that the PSR is located beyond the tangent point (8.4 kpc) but
not significantly beyond the Solar circle (19 kpc).
The electron density model by \cite{Lyne85} predicted a distance of 
14.5 kpc to B1849+00. 
The new electron density model by Cordes \&
Lazio (2002) places the PSR at a dispersion measure distance of
$(8.4 \pm 1.7)$ kpc, while 
the most recent model for the Galactic rotation by \cite{Fich89}
places a HI kinematic lower limit to its distance of 7 kpc.

PSR B1849+00 is located eight arcmin south of
the center of the SNR G33.6+0.1 (also known as Kes 79, 4C00.70, HC13). 
Two extragalactic point sources, 1849+005 and 1850+009 are located 14 arcmin 
south-west and 14 arcmin north, respectively, of G33.6+0.1.
G33.6+0.1 is a shell type SNR with diameter of about 10 arcmin
\citep{Caswell81}. 21-cm continuum observations by \cite{Frail89} 
confirmed its morphology and a slightly larger diameter of $\sim$16 arcmin.  
HI absorption measurements towards B1849+00, G33.6+0.1, 1849+005 and
1850+009 by \cite{Frail89} concluded that 
1849+005 and 1850+009 are indeed extragalactic sources, and that the
SNR lies in front of the PSR, at the distance of $(10\pm2)$ kpc. 
This distance was determined by scaling the PSR distance of 14.5 kpc
by the ratio of HI optical depth integrals of the SNR and the PSR.
We note though that this method is very uncertain especially at low
Galactic latitudes
where any change in optical depth profiles along nearby lines-of-sight
is more likely to be due to density or temperature variations in 
the absorbing cloud then due to one of the background objects being more distant.

Several studies have investigated a possible relationship between 
B1849+00 and G33.6+0.1.
\cite{Han97} suggested that the two objects are related and that B1849+00 was
born inside G33.6+0.1. However, very
recently, \cite{Seward02} detected a compact object with
Chandra, right in the center of G33.6+0.1, which seems to be a much stronger
candidate for the neutron star created in the SNR explosion. 

The unusually high dispersion measure of B1849+00 and its proximity to
G33.6+0.1 motivated several authors to investigate whether the SNR is
(partially) responsible for the interstellar scattering along the PSR's
line-of-sight. In particular, \cite{Spangler86} investigated whether
the turbulence toward the PSR could be driven by diffuse shock 
acceleration upstream of
the SNR boundary. The uncertain distances to the two objects hindered
conclusive results. \cite{Frail89} were also unable to distinguish
whether the scattering occurs from a thin screen provided by the SNR or
from a more general type of Galactic turbulence.

There were several previous investigations of molecular gas
in the direction of G33.6+0.1 \citep{Turner79,Scoville87,Green89,Green92}. 
OH absorption and emission features were observed by Turner (1979) and
Green (1989). $^{12}$CO(1-0) observations by Scoville et al. (1979) found 
a molecular cloud in the same direction. 
Green \& Dewdney (1992) observed bright and extended HCO$^{+}$ emission  
in the eastern portion of the SNR at velocities corresponding to the
molecular cloud and concluded that the SNR is most likely interacting with
the cloud.
\cite{Green97} detected emission at 1720 MHz associated with the
SNR using the Parkes telescope, however compact maser emission was not detected
with the VLA observations by \cite{Koralesky98}.

\section{Observations and data processing}
\label{s:observations}

\subsection{Arecibo OH Observations}
OH observations were undertaken in early January 2000 toward seven pulsars,
using the Arecibo 305-m telescope\footnote{The Arecibo Observatory is 
part of the National Astronomy
and Ionosphere Center, operated by Cornell University under a
cooperative agreement with the National Science Foundation.}. 
The observed pulsars were:
B1737+13, B1849+00, B1933+16, B2016+28, B1944+17, B1915+13 and B1929+10.
These pulsars were selected from the Princeton Pulsar Catalogue
\citep{Taylor93} as having low galactic latitude,
a reasonably high flux density at
21-cm and previously detected sharp HI absorption features.
The OH absorption was detected only against one of our observed sources -- 
PSR B1849+00. A marginal detection was found in direction of B1915+13 which
requires further confirmation.

\subsubsection{Observing Technique}

The Gregorian feed was used with the ``L-band wide'' receiver 
(with frequency range 
1.12 -- 1.73 GHz). The illuminated part of the 305-m dish covers an 
area of about 210$\times$240 m, resulting in a beam FWHM of approximately
$2'.6\times3'.0$ at 1.6 GHz.
The Caltech Baseband Recorder (CBR; \cite{Jenet97}) was used as a 
fast-sampling backend, with a total 
bandwidth of 10 MHz at each of two orthogonal circular polarizations, 
simultaneously covering both OH mainlines at 1665.4018 and 1667.3590 MHz.
The raw complex voltage data samples
were recorded every 100 ns and stored directly on DLT tapes for further analysis.
Observations were undertaken in blocks of `ON'-frequency scans, centered
on 1664.4 MHz and having duration of about 30 min, and  
`OFF'-frequency scans, centered at 1666.4 MHz, 
required to perform frequency switching to flatten baselines, lasting about 5 min.
The observing frequency was corrected for the changing Doppler shift 
of the observatory with respect to the local standard of rest (LSR) 
at each frequency change. 
In particular, PSR B1849+00 is observable at Arecibo only for about one hour
each day, and at the zenith angle of about 18 degrees. The system
temperature at this zenith angle was about 40 K.
In total, 2.8 hours were spent for this object on the `ON'-frequency scans,
and about 40 min on the `OFF'-frequency scans.
All data tapes were sent to Caltech's Center for Advance
Computation and Research (CACR) for further processing.

\subsubsection{Off-line data processing}
\label{s:off-line-processing}

The fast--sampled raw data were Fourier--transformed to provide a sequence
of dedispersed spectra, each consisting of 2048 frequency channels across 
a 10 MHz total bandwidth for the two orthogonal polarizations.  
(See \cite{Jenet97} for details of the algorithm.)  
The spectra have frequency resolution of 4.9 kHz, or velocity
resolution of 0.9 \kms, without any smoothing.
These spectra were then accumulated
modulo the apparent pulsar period into one of 128 pulse--rotational--phase 
bins.  The five-- to thirty--minute scan is now represented by a data cube of 
temperature as a function of frequency and pulse phase.

The next stage of data processing involved extraction of `ON'-pulse and
`OFF'-pulse spectra from the 3-dimensional data cube for each scan. 
Our software  finds the pulsar pulse in the cube, and weights (by
${T_{\rm PSR}}^2$) and accumulates 
those spectra gathered during the pulse into a raw ``pulsar--on'' spectrum 
and those measured between pulses into a ``pulsar--off'' spectrum.
Two types of spectra of astrophysical interest are then formed.   
The grand pulsar absorption spectrum (see Fig. 1),
which depicts the pulsar signal alone
(as absorbed by any intervening OH), takes advantage of the pulsed nature
of pulsar radiation.  The  grand pulsar absorption spectrum is created by 
generating the
``pulsar--on'' -- ``pulsar--off'' spectrum for a scan;  then doing frequency 
switching to flatten the baseline; and finally accumulating all such spectra
with a weight again proportional to ${T_{\rm PSR}}^2$.  The  grand ``pulsar--off''
spectrum (see Fig. 2), which registers all emission and absorption lying in the 
telescope beam during the time that the pulsar signal is not present,  is 
created by summing all scans' ``pulsar--off'' spectra, and then fitting 
and removing a polynomial to flatten the baseline.  
(We decided to fit a polynomial to the grand ``pulsar-off'' spectrum, 
rather than doing frequency switching, 
because of the great complexity of the spectrum; multiple 
emission and absorption features were present at both OH
mainlines, with the line separation of 1.95 MHz and the frequency
switching offset of 2 MHz.)

The normalized grand pulsar absorption spectra, at 1665 and 1667 MHz, 
are in units of $I(v) / I_{0}=e^{-\tau(v)}$, 
the optical depth of the intervening OH, and do not require any
further calibration. 
The grand ``pulsar-off'' spectra are in units of brightness temperature and
need to be calibrated.
As sources of known temperature were not observed for the flux calibration
purpose, we calibrated these spectra by comparing them with previous
OH observations towards G33.6+0.1, obtained by \cite{Turner79} using the 
NRAO 140-ft telescope (FWHM of 18.8 arcmin).
A constant, overall, conversion factors were estimated  for both 1665 and 
1667 MHz lines to scale our spectra to match those presented by Turner
(1979). The spectra 
were then scaled down by a factor of 1.6 as the PSR is located
south of the SNR where continuum emission is weaker, therefore the
brightness temperature covered by the Arecibo beam is lower than the
brightness temperature at the center of the SNR. 
This additional scaling factor was estimated from HI contours presented in
Fig. 1 of \cite{Frail89}.
The rms noise in the resultant spectra agrees within 25\% with the
expected theoretical noise level calculated for the given integration time, 
bandwidth and the system temperature (including the SNR contribution). 
The optical depth was calculated using:
\begin{equation}
\tau(v)=- \ln \left (\frac{T_{\rm B}(v)}{T_{\rm B}^{\rm cont}} +1  \right )
\end{equation}
where brightness temperature of continuum emission $T_{\rm B}^{\rm cont}
\sim4$ K was assumed.

\subsection{12-m CO Observations}

G33.6+0.1 was previously  mapped in $^{12}$CO(1-0) with 
the 12-m telescope by \cite{Green89}, however
these data are not easily accessible in electronic form. We
re-observed a 25$'\times15'$ region around G33.6+0.1.
Observations were obtained in December 2002. Absolute on-the-fly mapping
was performed with a scanning rate of 60 $''$/sec and 
a reference position being RA 19$^{\rm h}$ 00$^{\rm m}$, Dec  $-01^{\circ}$
30$'$ (B1950). 
The planet Mars was used for pointing and focus. 
Two filter banks and the digital millimeter autocorrelator 
were used simultaneously to record data for two orthogonal polarizations, 
resulting in final velocity resolutions
of 1.3, 2.6 and 0.1 \kms~per channel. 
The data reduction was performed in the {\sc aips} package. 
Line-free channels were used for continuum
subtraction. Data were gridded using the {\sc sdgrd} task.
The rms noise level is 0.4 K per 1.3 \kms~wide channels.
At 115 GHz the telescope FWHM is 55 arcsec.
Main beam efficiency is 0.62 and forward spillover and efficiency is 0.72. To
convert from recorded $T_{\rm R}$ to brightness temperature $T_{\rm B}$ 
multiplication by 1.16 is required. To further convert to flux density
multiplication by the antenna gain, 33 Jy K$^{-1}$, is required \citep{Helfer03}.

\section{Pulsar OH absorption spectrum toward PSR B1849+00}
\label{s:psr-only}

\begin{figure*}
\vspace{0.5cm}
\plotone{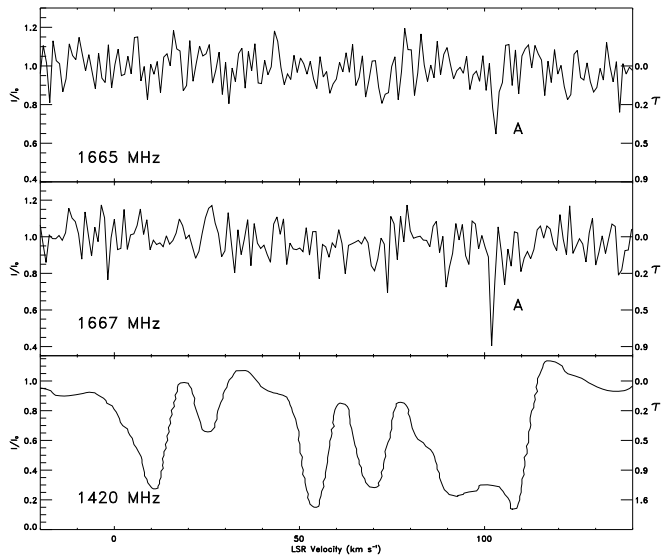}
\vspace{0.5cm}
\caption{\label{f:psr_abs}Pulsar absorption spectra toward
B1849+00 at 1665 and 1667 MHz (first and second panel). 
The 1420 MHz absorption spectrum against the PSR from Clifton et al. (1988)
(third panel). Each spectrum is a difference spectrum between 
``pulsar-on'' and ``pulsar-off'' spectra, and hence shows 
absorption produced against the PSR only.}
\end{figure*}

Fig.~\ref{f:psr_abs} (first and second panel) shows the pulsar OH absorption
spectra, at 1665 and 1667 MHz, toward B1849+00. 
At both frequencies, narrow absorption lines were detected at velocity of
about 102 \kms. This is the first successful OH absorption, 
to our knowledge, in both OH mainlines detected toward any pulsar.
The optical depth noise level is 0.08.
The detected absorption lines have an intensity of $>5$-$\sigma$.

As described above in Section~\ref{s:off-line-processing}, 
these spectra depict the   pulsar signal 
{\it{alone}} being absorbed  by intervening OH.  It is important to note
that emission and absorption due to any other source, astrophysical or 
instrumental, has been eliminated via the differencing procedure that 
generated these spectra.
The units on the left-hand axis on Fig.~\ref{f:psr_abs} are for 
the normalized pulsar intensity ($I/I_{0}$),
while the right-hand axis shows the optical depth of OH lying between 
the pulsar and Earth ($\tau=- \ln(I/I_{0})$). 
Several additional weak absorption features could be present in the OH
spectra presented in Fig.~\ref{f:psr_abs}. In particular, there may be a 
hint of an absorption feature seen at 1667 MHz around velocity of 75
\kms~where a strong CO emission is found
(see Fig.~\ref{f:psr_off}). However, we concentrate in this paper only on 
more than 3-$\sigma$ detections.
The last panel in Fig.~\ref{f:psr_abs} shows the HI absorption of signals 
from PSR B1849+00, adopted from Clifton et al. (1988).

The absorption system shown in Fig.~\ref{f:psr_abs}, which we label as `A',
seems to have an unusually high optical depth. 
Typical OH absorption found in the surveys against extra-galactic sources
has $\langle \tau_{1667} \rangle \sim 0.05$ and
$\langle {\rm FWHM} \rangle \sim 1$
\kms~\citep{Dickey81,Colgan89,Liszt96}. Higher optical depths are quite
rare and are found close to SNRs and/or HII regions 
\citep{Goss68,Manchester71,Yusef02}.

We fitted Gaussian functions to the PSR optical depth profiles, 
$\tau=\tau_{\rm max}\exp[-4 \ln2 (v-v_{0})^{2}/{\rm FWHM}^{2}]$ and 
Table 1 lists the derived parameters. Columns 3 -- 5 give $\tau_{\rm max}$, 
$v_{0}$ and FWHM, which are the peak height, velocity center and full width
at half maximum, respectively. Column 6 gives the equivalent width, ${\rm EW}= \int
\tau dv$, and column 7 gives the ratio of equivalent widths for 1667 and
1665 MHz lines, $R_{\tau}={\rm EW(1667)}/{\rm EW(1665)}$.
$R_{\tau}=1.8$ for thermalized level populations of any optical depth \citep{Dickey81}.
Column 8 gives the ratio of OH column 
density $N_{\rm OH}$ to the excitation temperature $T_{\rm ex}$, determined by:
\begin{equation}
\frac{N_{\rm OH}}{T_{\rm ex}}= \frac{C_{0}}{f} \int \tau dv
\end{equation}
where $C_{0}=4.0\times10^{14}$, for the 1665 MHz line, and
$C_{0}=2.24\times10^{14}$ cm$^{-2}$ K$^{-1}$ (\kms)$^{-1}$ for the 1667 MHz
line. The OH solid angle filling factor $f$
is assumed to be unity for these calculations.

Absorption features seen in Fig.~\ref{f:psr_abs} (see also Table 1) are 
very narrow with the velocity FWHM of 1.5 and 1.1 \kms, respectively.
In addition, central velocities of feature `A' at 1665 and 1667 MHz
differ by about 1 \kms. This is very surprising result. There is no
instrumental or processing step, to our knowledge, that could cause such
an offset. The two OH main line rest frequencies
are known to 100 Hz accuracy \citep{Meulen72}, while the offset
between central frequencies of feature `A' at 1665 and 1667 MHz is 5.6 kHz.
The 1665 and 1667 MHz spectra are actually two chunks
of a single 10 MHz wide spectrum, making an instrumental channel offset problem
unlikely.
It is also hard to explain this offset as being due to an astronomical reason.
The only possible reason we can think of are shock effects that could
somehow cause  population difference in the lower states of the two OH main
transitions. However, we are not aware of theoretical models of this phenomenon.
Certainly, further observations with higher velocity resolution are 
necessary to confirm this offset.

{\small
\begin{table*}
\caption{\label{t:clumps_table}Fitted OH parameters.}
\begin{tabular}{lccccccc}
\noalign{\smallskip} \hline \hline \noalign{\smallskip}
Feature & Transition & $\tau_{\rm max}$ & $v_{0}$ & ${\rm FWHM}_{v}$ &
EW & $R_{\tau}$ & $\frac{N_{\rm OH}}{T_{\rm ex}}$\\
 & (MHz) &  & (\kms) & (\kms) & (\kms) &  & $10^{14}$ (cm$^{-2}$ K$^{-1}$)\\
\noalign{\smallskip} \hline \noalign{\smallskip}
{\bf Pulsar} & & & & & &\\

A&   1665& $0.4\pm0.1$    &$103.0\pm0.2$ &$1.5\pm0.4$ &0.8  &1.5  & 3.2\\
 &   1667& $0.9\pm0.1$ & $102.0\pm0.1$&$1.1\pm0.2$ &1.2  &     & 2.7 \\

{\bf SNR} & & & & & & &\\

A& 1665& $0.010\pm0.002$& $101.6\pm0.9$&$11.9\pm1.3$&0.14 & & 1.7\\
 & 1665& $0.049\pm0.002$& $104.4\pm0.1$& $4.7\pm0.2$&0.28 & & \\

B & 1665& $0.003\pm0.001$&$8.7\pm2.2$ &$10.4\pm3.8$&0.04&1.0 &0.5 \\
  & 1665&$0.021\pm0.002$&$11.9\pm0.1$&$3.0\pm0.3$&0.08&2.7 & \\

\hline
A& 1667& $0.016\pm0.003$& $94.0\pm0.4$& $4.7\pm0.9$&0.09 & & 0.8\\
 & 1667& $0.021\pm0.001$& $101.8\pm0.5$& $9.8\pm1.3$&0.25 & & \\

B& 1667& $0.007\pm0.003$ &$9\pm3$&$5\pm4$ &0.04& &0.6 \\
 & 1667& $0.066\pm0.009$ &$11.7\pm0.1$ &$2.8\pm0.2$ &0.22& & \\

C& 1667&$-0.012\pm0.003$&$70.1\pm1.3$&$6.6\pm1.6$&$-0.1$ & & \\
 & 1667&$-0.030\pm0.006$&$72.6\pm0.1$&$2.7\pm0.5$&$-0.1$ & & \\

\noalign{\smallskip} \hline \noalign{\smallskip}
\end{tabular}
\end{table*}
}

\section{Pulsar OFF Spectrum}
\label{s:snr-only}

\begin{figure*}
\plotone{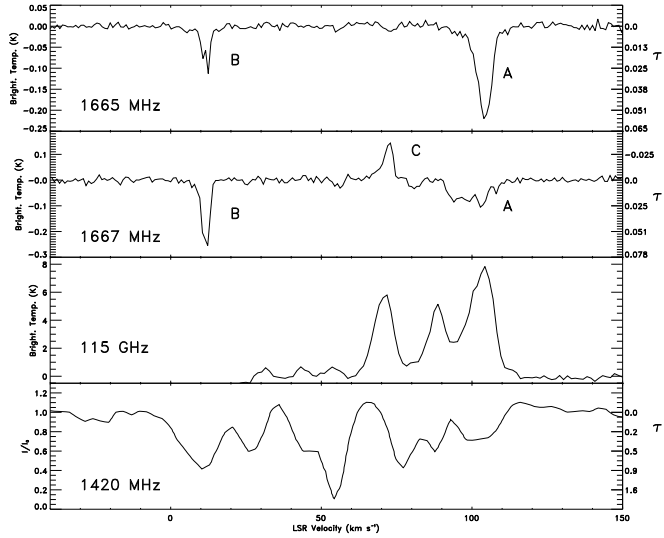}
\vspace{1cm}
\caption{\label{f:psr_off}The ``pulsar-off'' spectrum toward B1849+00 at 1665
(first panel) and 1667 (second panel). Absorption and emission
features seen in this spectrum, and labeled
as `A', `B' and `C', are observed against SNR B33.6+0.1 which is partially covered
by the Arecibo beam. 
The $^{12}$CO(1-0) emission spectra, obtained with the 12-m telescope 
in direction toward the PSR and averaged over the whole Arecibo beam, 
$\sim$ 3$'$, (third panel).
The 1420 MHz spectrum 
(fourth panel), is the HI absorption spectrum toward the SNR,
from Frail \& Clifton (1989). }
\end{figure*}

\begin{figure*}
\epsscale{1.0}
\plotone{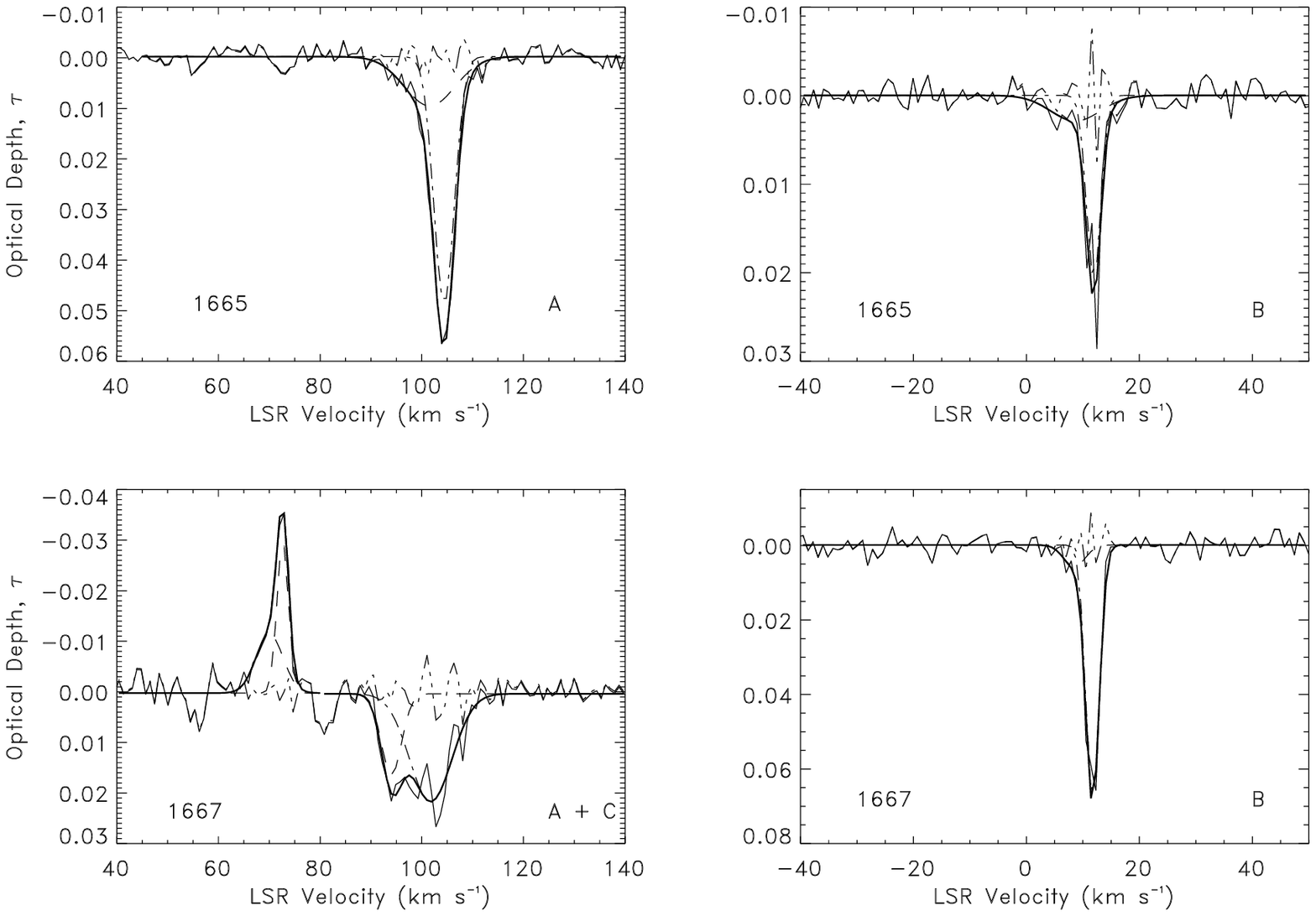}
\vspace{0.5cm}
\caption{\label{f:spectra_fits}
Gaussian fits to the OH spectral lines, observed in Fig.~\ref{f:psr_off}, 
against the SNR (thin solid line):
wide Gaussian component (dashed line), narrow Gaussian 
component (dot-dashed line), 
final fit as a sum of the two components (thick solid line), and 
residuals (dot-dot-dashed line).}
\end{figure*}

Fig.~\ref{f:psr_off} (first and second panel) shows the ``pulsar-off'' 
spectrum 
(see Section~\ref{s:off-line-processing}). As the SNR G33.6+0.1 is
centered only 8 arcmin from B1849+00 and its diameter is $\sim$ 16 arcmin
(from Frail \& Clifton 1989, their Fig. 1),
while the Arecibo telescope FWHM at 1666 MHz is $2.6' \times 3.0'$, the 
absorption
features in this spectrum are effectively produced by intervening OH lying
in front of G33.6+0.1. 
The third panel on Fig.~\ref{f:psr_off} shows $^{12}$CO(1-0) emission in the
direction of B1849+00 integrated over the Arecibo beam. The forth panel in
Fig.~\ref{f:psr_off} shows the 1420 MHz absorption spectrum toward G33.6+0.1 
from Frail \& Clifton (1989).

The ``pulsar-off'' spectra show two absorption systems, centered at 
about 102 and 10 \kms, observed in both OH mainlines. 
The first one, `A', has been already seen in the PSR absorption
spectrum. We label the second one as `B'.
In addition, there is an emission feature, `C', seen 
at about 70 \kms. It is immediately obvious that these absorption
features differ greatly, in both peak intensity and linewidth, from
features seen in the pulsar absorption spectra in Figure~\ref{f:psr_abs}.
In particular, feature `A' in the 1667 MHz line is almost 15 times
wider and 30 times shallower that its
corresponding feature in the PSR absorption spectrum.

We fitted Gaussian functions to optical depth profiles of features 
`A', `B' and `C' and 
results are shown in Table 1. In all cases at least two Gaussian 
functions were required as observed profiles are broad. 
Usually, a wider Gaussian component is necessary to fit line wings, 
where optical depth decreases  more gradually.
The fitted Gaussian components and their corresponding residuals are shown in 
Fig.~\ref{f:spectra_fits}.
Absolute values for the peak optical depth are at least 10
times lower than for the pulsar absorption spectra, but are similar to
values obtained in several OH surveys toward extragalactic sources 
\citep{Dickey81,Colgan89,Liszt96}. 
The FWHM of narrow Gaussian components is typically 3-5 \kms, while for the
wider component is $\sim10$ \kms.

Wide absorption/emission profiles could be caused by blending of 
molecular clumps
along the line of sight \citep{Marscher91,Marscher94}. 
Another possible cause of wide absorption profiles is 
propagation of a C-type shock through a molecular
cloud, which could occur in the case of a cloud-SNR interaction.
\cite{Flower98} modeled OH absorption line profiles in the presence
of C-type shocks and found double peaks arising from contributions from
both the host (quiescent) and shocked gas. Their double components have
comparable peak values but the quiescent one is significantly narrower that the
shocked one. Fig.~\ref{f:spectra_fits} and Table 1 show that feature `A'
at 1667 MHz is reminiscent of the C-type model profiles.
This particular OH absorption line was interpreted previously as a signature of an
interaction between the SNR and a molecular cloud \citep{Green89,Green92}.
However, in three other cases the wider Gaussian component is associated 
with the line wing only 
and does not resemble profiles expected from  C-type shocks.

In the case of feature `A' wide Gaussian components at both 1665 and 1667
MHz agree well in central velocity and FWHM, we then use their equivalent
widths to estimate $R_{\tau}=1.5$. Narrow components have very different
central velocity. Similarly, for feature `B' narrow Gaussian components
agree well in central velocity and FWHM, and have  $R_{\tau}=2.7$. Wider
Gaussian components for feature `B' have the same central velocity but
significantly different FWHM, in this case $R_{\tau}=1.0$.

\section{CO distribution}
\label{s:CO}

\begin{figure*}
\epsscale{1.0}
\plotone{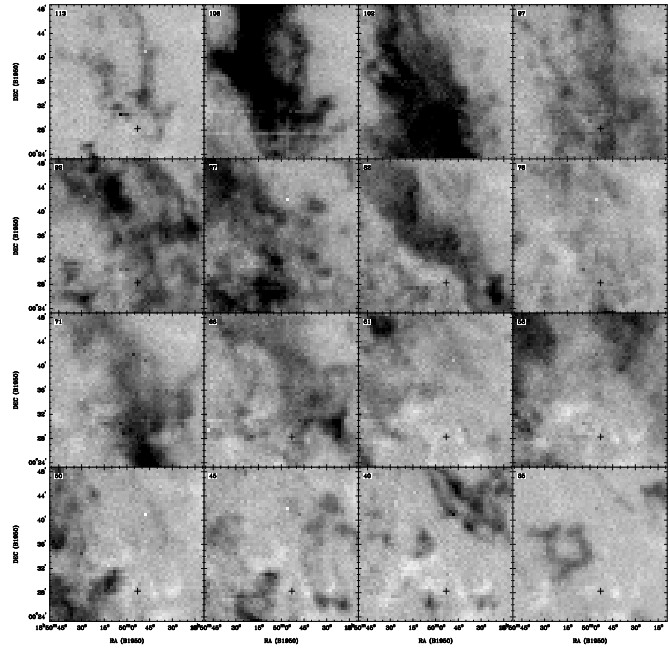}
\vspace{1cm}
\caption{\label{f:CO_lowres} 
Right ascension-declination images of $^{12}$CO(1-0) emission 
at different LSR velocities, given in the
top left corner of each panel, obtained with the 12-m telescope. 
The grey-scale range is $-2.3$ to 7.5 K, with a linear
transfer function. Position of B1849+00 is given with a cross.}
\end{figure*}

\begin{figure*}
\plotone{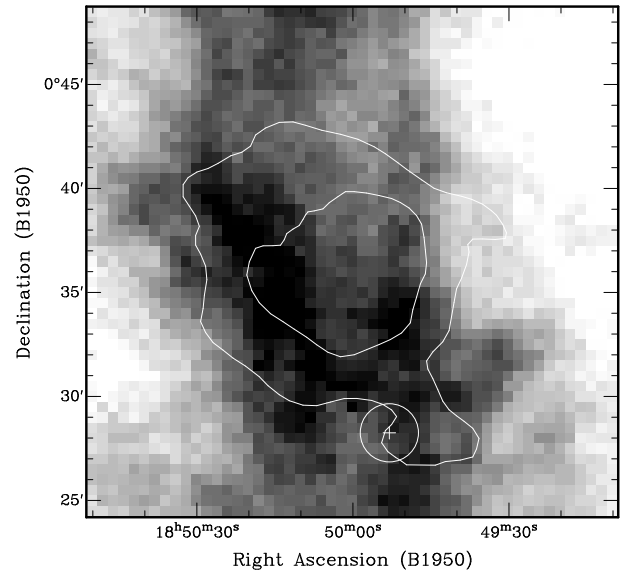}
\caption{\label{f:g33_composed} 
Integrated $^{12}$CO(1-0) intensity image obtained with the 12-m
telescope along with 21-cm continuum contours of the SNR
from the VLA. Integrated velocity range covers 95 to 113 \kms. The overlaid
contours are from the the 21-cm continuum image of
G33.6+0.1 from Frail \& Clifton (1989), at 2 and 16 K levels of 
brightness temperature. The PSR position as shown with a cross and the
Arecibo beam is shown with a circle (FWHM of $\sim$3 arcmin).}
\end{figure*}

Fig.~\ref{f:CO_lowres} shows $^{12}$CO(1-0) emission at different LSR
velocities. The CO data
presented here complement data published by \cite{Green92} 
by covering a wider velocity
range. A large molecular cloud (with size of $>25$ arcmin)  
is seen between 97 and 113 \kms~that extends over
the whole continuum distribution of G33.6+0.1, see $^{12}$CO(1-0) integrated
intensity image in Fig.~\ref{f:g33_composed}. 
This cloud was first cataloged by \cite{Scoville87} who
estimated its average linear size of $\sim$40 pc and a
mean molecular hydrogen density of $\sim$180 cm$^{-3}$. 
The CO emission is particularly strong
on the west part of the SNR. At several positions along this 
side \cite{Green92} also found strong HCO$^{+}$ emission which was 
interpreted as being due to an interaction between the SNR and the molecular cloud.
The cloud is elongated along the Galactic plane (in the direction 
from the north-west to the south-east) and has somewhat sharper 
edge on the lower latitude side.

An interesting loop-like feature is seen around 113 \kms, extended along
the Galactic plane.
Around 92 \kms~a higher latitude cloud is seen that transforms into 
a long filament running parallel to the plane at about 82 \kms.
Although these are low sensitivity observations a wealth of small scale
molecular features is found, particularly around 92 and 40--42 \kms.

Two small clouds, probably belonging to larger complexes, are seen at
56 \kms, RA 18$^{\rm h}$ 49$^{\rm m}$ 40$^{\rm s}$,
Dec 00$^{\circ}$ 40$'$ 00$''$, and at $\sim$48--50 \kms,
RA 18$^{\rm h}$ 50$^{\rm m}$ 08$^{\rm s}$ Dec 00$^{\circ}$ 30$'$ 0.14$''$.
The peak brightness temperature of clouds is 7.5 and 8 K and they both 
lie in the direction of the continuum emission associated with G33.6+0.1.
Interestingly, Turner (1979) observed an OH absorption feature at 57
\kms~in 1667 MHz line, accompanied with an emission feature at 60
\kms~in 1665 MHz lines. Observations by \cite{Green89} at 1667 MHz 
showed a flip from an absorption at 55 \kms~to an emission at 60 \kms. 
An interesting donut-like  feature is seen around 35 \kms.

Figure~\ref{f:g33_composed} shows the integrated $^{12}$CO(1-0) intensity
distribution over velocity range 95 to 113 \kms, near our feature `A'.
Contours of the
21-cm continuum image of G33.6+0.1 by Frail \& Clifton (1989) are also
overlaid. The mean velocity of the cloud is 106 \kms, no velocity gradient
is found.

\section{Discussion}
\label{s:small-scale-structure}

\begin{figure*}
\plotone{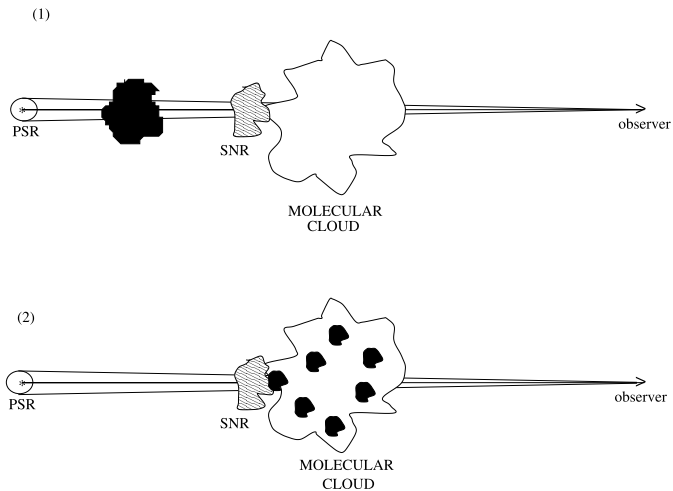}
\caption{\label{f:scenarios}
Two possible scenarios for the origin of OH absorption lines.
(1) An additional molecular cloud (shown in black)
is located in front of PSR but behind the SNR producing
the PSR absorption spectra (shown in Fig.~\ref{f:psr_abs}).
The PSR absorption spectra measure absorption only in the tiny column in front of
the PSR (illustrated as a line), while the ``pulsar-off'' spectra (shown in
Fig.~\ref{f:psr_off}) measure absorption (and emission) throughout 
the illustrated cone that covers a portion of the SNR (FWHM $\sim$ 2 arcmin).
(2) All molecular gas is in front of both the SNR and the PSR.
The PSR absorption features are produced by one
small clump (`cloudlet')
while the shallower, broader absorption against the SNR is produced by an ensemble
of small clumps (shown in black).}
\end{figure*}

What is particularly striking about the OH absorption data 
presented in this paper is that the difference in the observed optical depths 
(line shape, width and peak) in the PSR absorption spectrum
(Fig.~\ref{f:psr_abs}) and the ``pulsar-off'' spectrum
(Fig.~\ref{f:psr_off}, 1st and 2nd panel) is so great.
 The PSR absorption spectrum shows absorption by the
intervening OH against the pulsar continuum emission {\it alone}, while the  
``pulsar-off'' spectrum shows absorption produced against 
the continuum emission from G33.6+0.1 {\it alone} within the Arecibo beam.  
In other words, spectra in Figs.~\ref{f:psr_abs} and~\ref{f:psr_off} show
two different absorption features along {\it the same} line-of-sight and
with the same central velocity around 103 \kms.
In particular, the absorption features seen in the PSR absorption
spectra at 1665 and 1667 MHz
are deep  and narrow, but with the ratio of equivalent widths for 1665 and 1667 MHz
lines being very close to 5:9. 
The line ratio suggests that the observed
absorption spectra are not the result of an instrumental or processing artefact, 
but that originate from molecular gas along the 
line of sight toward the PSR.

From Table 1, $N_{\rm OH}/T_{\rm ex}\sim 3 \times10^{14}$ cm$^{-2}$
K$^{-1}$ for feature `A' in the PSR absorption spectra.
It is typically assumed that $T_{\rm ex} \sim$5 K 
for clouds unassociated with HII regions, and around 10 K for clouds
associated with HII regions \citep{Bourke01}. Liszt \& Lucas (1996)
have found that gas seen in OH absorption has typically  $T_{\rm ex} \sim$4
K, while gas seen in OH emission is warmer with $T_{\rm ex} \sim$7--13 K.
With assumption of $T_{\rm ex} \sim$5--10 K,
we estimate $N_{\rm OH} \sim {\rm few} \times 10^{15}$ cm$^{-2}$.  
Under assumption that column densities of OH and hydrogen (HI$+$H$_{2}$)
are related with $N_{\rm H} \sim 10^{8}N_{\rm OH}$, we get
$N_{\rm H} \sim {\rm few} \times 10^{23}$ cm$^{-2}$ for this feature.
This is higher than what is usually sampled by OH \citep{Bourke01}.

In Section~\ref{s:CO} we showed that a large molecular
cloud is seen in the SNR and the PSR direction. The line shapes (see 
Section~\ref{s:snr-only}) and previous work by 
\cite{Green89} and \cite{Green92} suggest that the SNR is most likely 
interacting with the molecular cloud. To explain the large difference in 
absorption optical depths against the PSR and SNR
we investigate two possible geometrical scenarios.
 
\subsection{An additional molecular cloud is located in front of the PSR
yet behind the SNR}

It is possible that
absorption features seen in the PSR absorption spectrum and the
``pulsar-off'' spectrum are not related physically. This could
happen in the case where the sharp and deep absorption lines,
seen in the PSR absorption spectra, come from an additional molecular 
cloud located in front of the PSR yet behind the SNR (see Fig.~\ref{f:scenarios} 
case 1 for a graphical representation). 
The broad OH absorption lines in the ``pulsar-off'' spectrum are, most likely, a 
result of the interaction between the SNR and the molecular cloud, 
while the narrow absorption lines are not associated with the cloud-SNR
interaction. This is similar to different 
OH absorption lines observed in direction to W28 \citep{Yusef02}.

This additional molecular cloud could be of any size.
The large-scale $^{12}$CO(1-0) images in Fig.~\ref{f:CO_lowres} and
Fig.~\ref{f:g33_composed} do not show any obvious structure that
could be associated with this secondary cloud. 
To investigate this scenario we compared the hydrogen column density derived
from OH and CO in the PSR direction. The 3rd panel in 
Fig.~\ref{f:psr_off} shows the CO emission
spectrum which has several components, the last one
being centered at 103.2 \kms. This component has the integrated CO
intensity of 75.9 K \kms. Using the conversion factor between the
integrated CO intensity and H$_{2}$ column density by \cite{Dame01} we
derive $N({\rm H_{2}})=1.5\times10^{23}$ cm$^{-2}$ in this direction, or
$N_{\rm H}=3\times10^{23}$ cm$^{-2}$. 
This agrees well
with $N_{\rm H}=2-4\times 10^{23}$ cm$^{-2}$ derived from the total OH
column density traced by the PSR absorption spectrum and the ``pulsar-off''
spectrum. This suggests that, most likely, all OH seen in
absorption and CO seen in emission coexist in
the same region and that the existence of an additional molecular cloud along
the line of sight is not very likely explanation.

\subsection{All molecular gas is in front of the SNR and the PSR}

Another possibility is that the OH absorption features seen in the PSR
absorption and ``pulsar-off'' spectra originate from {\it the same} 
general molecular cloud located in front of both the PSR and the SNR.
This could happen in the case where the PSR absorption is produced by a 
small clump (`cloudlet'), while the shallower, broader absorption 
features against the SNR are caused by an ensemble of `cloudlets' of 
varying properties (Fig.~\ref{f:scenarios} case 2).
Furthermore, the small `cloudlet' could represent a typical building 
block for the molecular cloud. 

What is particularly interesting about this scenario is that 
it would be a clear demonstration that a pencil-sharp 
absorption sample can differ {\it dramatically} from a large-angle 
absorption sample. 
This OH result would be very different from  HI absorption findings. 
\cite{Dickey79} and \cite{Payne82} observed continuum sources over a large 
range of solid angels and compared results
to those against small solid-angle sources. 
They found no obvious difference. \cite{Dickey81} observed 21-cm 
absorption in front of pulsars and extragalactic sources of 
widely varying angular sizes and 
concluded that the statistics of absorption are similar in 
all cases,  and that the 
`cloudlet' model of the interstellar HI is not prominent. However, 
the difference at HI and OH would not be totally unexpected:
the solid-angle effect is expected to be
more pronounced for molecular gas where clumpiness is known to be significant.

We explore some of the consequences of this scenario in the following two 
subsections.

\subsubsection{An estimate of the `cloudlet' size}

The solid angle subtended by the `cloudlet' intercepts
solid angles of both PSR and SNR continuum emission regions
(Fig.~\ref{f:g33_composed}). However, the ``pulsar-off'' spectrum 
does not appear to have a significant contribution from the
`cloudlet' seen in the PSR absorption spectrum. This suggests that 
the `cloudlet' covers a very small fraction of the SNR and 
can be used to place an upper limit on the `cloudlet' size. 

In this particular scenario, when all molecular gas is located in front of
both the PSR and the SNR, the observed optical depth profile against the SNR 
($\tau_{\rm snr}$, shown in Fig.~\ref{f:psr_off}) is a 
solid angle weighted average of the `cloudlet' optical depth profile, which
we will assume to be the same as the PSR optical depth profile 
($\tau_{\rm c}$, shown in Fig.~\ref{f:psr_abs}), and 
the pure optical depth profile against the SNR ($\tau'_{\rm snr}$) that
would be uncontaminated by the `cloudlet' absorption: 
\begin{equation}
\frac{\Omega'_{\rm snr} \tau'_{\rm snr} + \Omega_{\rm c} \tau_{\rm c}}
{\Omega'_{\rm snr} + \Omega_{\rm c}} = \tau_{\rm snr}.
\end{equation}
Here, $\Omega'_{\rm snr}$ and $\Omega_{\rm c}$ are solid angles of the 
SNR and PSR continuum emission occupied by OH seen in absorption. 
We assume that $\tau_{\rm snr}=k\tau'_{\rm snr}$, where
$k$ is determined from comparison of our OH observations with previous ones.
Previous OH observations of G33.6+0.1 were obtained with much larger
telescope beams covering the whole SNR (Turner 1979; Green 1989) and
therefore averaging optical depth profiles over larger solid angles. 
Hence these spectra can be assumed to be $\tau'_{\rm snr}$.
Our OH spectra agree with previous observations within 60\% (see 
Section~\ref{s:off-line-processing}), meaning that $k\geq1.6$. 
If we assume that $\Omega'_{\rm snr}$ is less or equal a half of the solid angle
subtended by the Arecibo beam (as suggested in Fig.~\ref{f:g33_composed}), 
then the peak optical depths listed in Table 1 allow us to place an estimate of
$\Omega_{\rm c}\leq0.25$ arcmin$^{2}$.

At the distance of 7 kpc $\Omega_{\rm c}$ translates to a maximum linear size of 1 pc
for the OH absorbing `cloudlet'. 
Using the values for $N_{\rm OH}/T_{\rm ex}$ from Table 1,  
we can now estimate the total column density of hydrogen in the
`cloudlet', and from there its volume density ($n$). Under assumption of a simple
spherical geometry, $n=N_{\rm OH}/(4\times 10^{10} \times r_{\rm pc})$, 
and $T_{\rm ex}\approx 10$ K, we estimate $n>10^{5}$ cm$^{-3}$. 
The hydrogen volume density and `cloudlet' size suggest that this is 
a very small and dense cloud of molecular gas,
similar to the tiny-scale molecular features 
on sizes of order of 10 AU discovered by \cite{Moore95}.

\subsubsection{Modeling optical depth profiles}

\begin{figure*}
\plotone{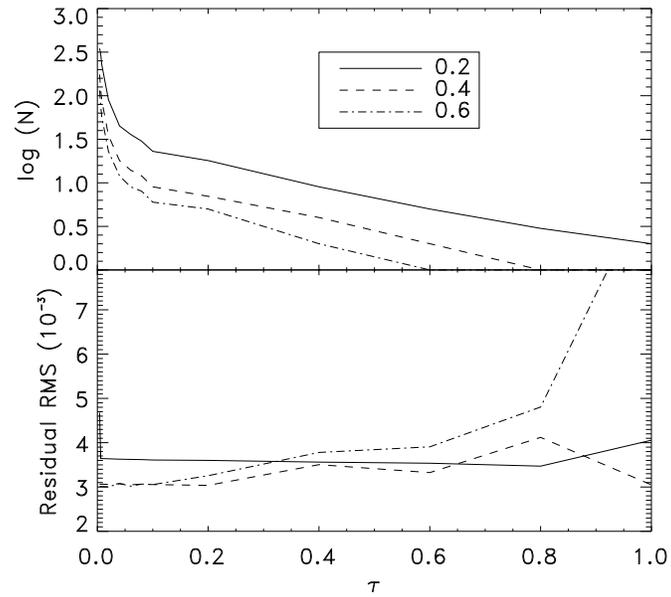}
\vspace{1cm}
\caption{\label{f:rms_nclump}Bottom:
rms of (modeled optical depth profile -- SNR optical depth profile at 1665 MHz)
as a function of peak optical depth of the template optical depth
profile ($\tau_{\rm max}$). Top: Natural logarithm of the number of 
template optical depth profiles
necessary to model the SNR optical depth profile at 1665 MHz as a function of
peak optical depth of the template optical depth profile.}
\end{figure*}

\begin{figure*}
\plotone{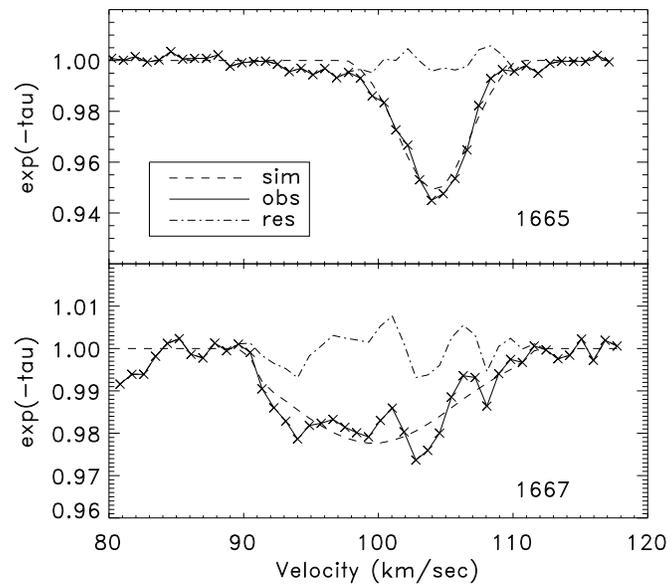}
\vspace{1cm}
\caption{\label{f:2plot_sim} The best modeled optical depth profiles 
obtained for the case
of $\sigma_{\rm c}=0.4$ \kms~and $\tau_{\rm max}=0.05$. Observed data are
shown with solid line, modeled with dashed line, and the residuals are
shown as dot-dashed line.}
\end{figure*}

We now investigate the possibility that the `cloudlet', seen in the PSR
absorption spectra, represents a
typical building block for the molecular cloud by attempting to model
broad absorption lines in the ``pulsar-off'' spectra with an ensemble of 
narrow absorption lines from the PSR absorption spectra.

We use a simple model for modeling optical depth profiles, 
based on the idea from Marscher, Bania \& Wang (1991). 
The model assumes a random distribution of a large number of `cloudlet' 
absorption features within boundaries of the molecular cloud. 
While `cloudlets', or basic building blocks, have thermal linewidths, the
observed final broad linewidths in the ``pulsar-off'' spectra are due 
to the bulk motion of individual `cloudlets'. We represent the `cloudlet' 
optical depth spectrum ($\tau_{\rm c}$) with a Gaussian
function, having the peak optical depth ($\tau_{\rm max}$), and the standard
deviation ($\sigma_{\rm c}$ in \kms).
`Cloudlet' optical depth profiles are centered along the line-of-sight
according to the probability distribution function.
We use a Gaussian form for this function which is determine
by fitting a single normalized Gaussian function to the ``pulsar-off'' 
optical depth profile. 

The velocity interval of interest, $\Delta V$, is then divided into 
$\Delta V/2 \sigma_c$ velocity bins, each of width $2 \sigma_c$.
The probability that each velocity bin contains 0, 1, 2, ... or $N$ `cloudlet'
optical depth profiles is given by the binomial statistics, and the 
expected value of e$^{-\tau}$, averaged over $\Delta V/2 \sigma_c$ velocity
bins, can be calculated using:
\begin{equation}
\langle e^{-\tau} \rangle = \frac{2\sigma_c}{\Delta V}
\sum_{i=1}^{\Delta V/2 \sigma_c} q_{i}^{N} 
\left [ \sum_{j=0}^{N} \left ( \begin{array}{c}
N \\ j
\end{array} \right ) \left ( \frac{p_{i}}{q_{i}} \right )^{j} e^{-j \tau_c} \right ]
\end{equation}
or
\begin{equation}
\langle e^{-\tau} \rangle = \frac{2\sigma_c}{\Delta V}
\sum_{i=1}^{\Delta V/2 \sigma_c} \left (p_{i}e^{-\tau_c}+q_{i} \right)^{N}.
\end{equation}
Here, $p_{i}$ is the probability that the central velocity of a `cloudlet'
belongs to the $i$th velocity bin, which is estimated from the probability
distribution function, while $q_{i}=(1-p_{i})$.
We have experimented with different velocity intervals $\Delta V$ 
and the best results are achieved for $\Delta V \sim 2.0\times$ FWHM of 
the single Gaussian fit to the ``pulsar-off'' optical depth profile. 

The simulations were performed for $\sigma_c=0.2$ \kms, the internal 
thermal velocity dispersion for gas with kinetic temperature of about 100 K.
This kinetic temperature was suggested by the HI 
absorption and emission data toward B1849+00 from Frail \& Clifton (1989).
In addition, we allowed a slight broadening of the internal `cloudlet'
velocity dispersion and performed simulations for $\sigma_c=0.4$ and 0.6 \kms. 
The last value is particularly of interest as it corresponds to 
the standard deviation of the PSR
absorption line from Table 1, while $\sigma_c=0.4$ \kms~is an intermediate
value between the pure thermal and the PSR linewidths. For each value of 
$\sigma_c$ and 
the range of $\tau_{\rm max}$ from 0.004 to 2.0, we estimated 
$\langle e^{-\tau} \rangle$ for various $N$ (ranging from 1 to 250). 
The simulated (expected) optical depth spectrum was then compared with the 
original SNR
spectrum, residuals were determined and the rms of residuals was
calculated. The best model fit was assumed to correspond to the lowest rms
residual value and the corresponding number of clumps, $N$, was taken as the
best estimate. For a given range of $\tau_{\rm max}$,
Fig.~\ref{f:rms_nclump} shows the rms of residuals
and the best number of clumps determined for the 1665 MHz SNR optical depth
profile, when $\sigma_c=0.2$ (solid line), 0.4 (dashed line)
and 0.6 \kms~(dot-dashed line).

The results of simulations show several general trends:
the rms of residuals slowly increases with $\tau_{\rm max}$; 
number of clumps needed to fit the SNR profile increases when $\tau_{\rm max}$
decreases; and number of clumps also increases when $\sigma_c$ decreases. 
In the case of $\sigma_c=0.4$ or 0.6 \kms, the simulated spectra are 
unsatisfactory (have high residual rms) for $\tau_{\rm max}>0.6$. 
In the case of $\sigma_c=0.2$ \kms~the rms stays quite stable all the way to
$\tau_{\rm max}\sim0.8$.
The lowest rms of residuals are achieved for $\tau_{\rm max}<0.1$
and $\sigma_c=0.4$ or 0.6 \kms.
Although good fits (small rms) are achieved for the wide parameter space
given by $\tau_{\rm max} \leq 0.6$ and any $\sigma_c$, the number of
required `cloudlets' is too small, less than 10, in most 
cases where $\tau_{\rm max}>0.1$.

For the case of $\sigma_c=0.4$ or 0.6 \kms, a large number of `cloudlets'
could make the molecular cloud only under very low peak optical depth conditions
of $\tau_{\rm max}\sim0.02-0.03$. The most meaningful results are achieved
for the thermal velocity dispersion, $\sigma_c=0.2$ \kms~with $\tau_{\rm
max}<0.1$, where $>30$ `cloudlets' are needed to model well 
the SNR optical depth profiles. 
Typical $\tau_{\rm max}$ obtained from surveys against extragalactic
sources \citep{Dickey81,Colgan89,Liszt96} is around 0.05.
As an example of how well optical depth profiles can be modeled, 
we show the case for $\sigma_c$=0.4 \kms~and $\tau_{\rm max}=0.05$ in
Figure~\ref{f:2plot_sim}. 
Although this is a very simple geometrical model it is possible to model
SNR spectra surprisingly well.
However, the 1665 MHz modeled profile is missing line wings, while the 1667 
MHz profile is extremely complex. If the thermal `cloudlet' linewidth 
is used, $N\sim50$ is required.

As a conclusion, the required number
of `cloudlets', with properties similar to the PSR optical depth profile, 
necessary for building the SNR profile is too small. The PSR profiles hence
can not be building blocks. Models of a large number of `cloudlets' having
optical depth profiles with thermal linewidth and $\tau_{\rm max}<0.1$
could work well.

\section{Summary}
\label{s:summary}

We have detected OH absorption against PSR B1849+00 at both 1665 and 1667
MHz OH mainlines using the Arecibo telescope. This is the first successful 
detection of OH in both mainlines toward any pulsar. 
The most important aspect of this detection is that it opens up a new
avenue for studying properties of molecular gas using extremely small solid angle
background sources. The OH absorption features detected against B1849+00
are unusually deep ($\tau=0.9$ at 1667 MHz) and narrow (FWHM$=$1.1
\kms). High OH optical depths are often found close to SNRs but with wide
linewidths that indicate an interaction between a SNR and a molecular cloud. 

In addition, we have detected several absorption and emission
features against G33.6+0.1, a nearby SNR that is partially
covered by the Arecibo beam. The most interesting result of this study is that
a pencil-sharp absorption sample probed by the PSR differs greatly from the
large-angle absorption sample probed by the SNR along the same line-of-sight.
In trying to understand this difference we have obtained $^{12}$CO(1-0)
observations of G33.6+0.1 with the 12-m telescope.

We have investigated two geometrical scenarios in order to explain the
observed OH absorption spectra. 
In the first scenario, an additional molecular cloud, located in
front of the PSR but behind the SNR, could be responsible for the
pencil-sharp absorption sample probed by the PSR.  
The main implication of the second scenario, whereby the OH absorption 
against the PSR and the SNR originate from the same molecular cloud, 
is that the statistics of OH absorption depend greatly on the size of the
background sources. This result is different from what was found for the
case of HI absorption studies, but not totally unexpected as clumpiness of
molecular gas is known to be significant.
We have further shown that in this scenario, the PSR is most likely probing a
very small molecular clump, `cloudlet', having angular size of $<30$ arcsec and
hydrogen volume density $> 10^{5}$ cm$^{-3}$. 

We have also
investigated the possibility that the `cloudlet' represents a typical building
block for the molecular cloud by modeling the SNR OH absorption profiles
with template `cloudlet' opacity profiles. It was shown that this is not the case,
however the SNR OH absorption profiles can be modeled well with template
profiles having peak opacity of $<$0.1 and velocity dispersion of
$\sim$ 0.2--0.4 \kms. High velocity and spatial resolution OH mapping of
the whole SNR is desirable to distinguish further between the two possible 
scenarios. In addition, further OH absorption measurements toward other pulsars are important
to constrain how common such phenomenon may be.

\acknowledgements
We would like to express our thanks to Caltech's Center for Advance
Computation and Research for the use of their facilities for data storage
and processing. We appreciate greatly Rick Jenet's help 
with processing of raw pulsar data.
S. S. is grateful to Alberto Bolatto and Jonathan Swift
for valuable input on the 12-m observing and data reduction strategy.
S. S. would also like to thank all staff members of the 12-m telescope for their
great support and help. We are grateful to an anonymous referee for helpful suggestions.
The National Radio Astronomy Observatory is a 
facility of the National Science Foundation operated under
cooperative agreement by Associated Universities, Inc.

\label{lastpage}
\end{document}